\documentclass{PoS}
\usepackage{subfigure}

\title{Nucleon-Nucleon Scattering and Large Nc QCD}

\ShortTitle{Nucleon-Nucleon Scattering and Large Nc QCD}

\author{\speaker{Boris A. Gelman}%
        \thanks{The work on which this talk is based was done in collaboration with T.D Cohen \cite{CohenGelman,NNobservables}.}\\
       New York City College of Technology, The City University of New York\\
       E-mail: \email{bgelman@citytech.cuny.edu}}

\abstract{
Nucleon-nucleon scattering observables are discussed in the context of large $N_c$ QCD.
As is well known, the baryon spectrum in the large Nc limit exhibits contracted $SU(2N_f)$
spin-flavor symmetry. This symmetry can be used to derive model-independent relations
between proton-proton and proton-neutron total cross sections. These relations are valid
in the kinematic regime in which the relative momentum of two nucleons is of order of $N_c$.
In this semiclassical regime the nucleon-nucleon scattering can be described in the
time-dependent mean field approximation. These model-independent results are compared
to experimental data for spin-independent and polarized total nucleon-nucleon cross sections.}

\FullConference{Xth Quark Confinement and the Hadron Spectrum,\\
		October 8-12, 2012\\
		TUM Campus Garching, Munich, Germany}

\begin{document}

\section{\label{sec:intro}Introduction}

Description of the nucleon-nucleon interaction, the basic
ingredient of nuclear physics, directly from quantum
chromodynamics (QCD) is a daunting task. The breakdown
of the pertubative expansion in terms of the QCD gauge coupling $g$
necessitates the use of alternative methods. One such method
proposed by 't Hooft in 1974 is to consider a QCD-like theory
with the number of colors $N_c$ and the gauge group $SU(N_c)$,
{\it large $N_c$ QCD} \cite{tHooft}.
The observables in large $N_c$ QCD are expanded in powers of $1/N_c$ around
the {\it large $N_c$ limit},
$g \rightarrow 0$, $N_c \rightarrow \infty$ and
finite $g^2 N_c$. In addition, it is {\it assumed} that large $N_c$
QCD is a confining theory and the asymptotic states are $SU(N_c)$
singlets. Despite our inability at present to evaluate even the leading
order terms, a great deal of insight comes from knowledge of the
scaling of hadronic observables in powers of $1/N_c$. The phenomenological
implications of large $N_c$ QCD are essentially  topological in nature.

The description of the meson and baryon observables in the large $N_C$ limit requires
different methods. Formally, it is due to the fact that the correlation
functions in the meson sector have a smooth expansion in powers of $1/N_c$
while the correlation functions in the baryon sector diverge in the large $N_c$ limit.
Physically, it is due to the fact that as shown by 't Hooft, the QCD in the large $N_c$ limit
is a theory of an infinite number of stable non-interacting mesons.\footnote{Additionally, the spectrum
of large $N_c$ QCD contains glueballs with vanishing mixing to mesons} The meson masses and sizes are
independent of $N_c$. This picture is phenomenologically satisfactory since in the real world mesons
interact weakly.

QCD is also a theory of strongly interacting baryons, the states carrying quantum numbers of the odd number of quarks. As was shown by Witten
\cite{Witten},
a consistent large $N_c$ description of strongly interacting baryons is possible. Remarkably, the same feature which on the one hand makes an analysis of the baryons in the large $N_c$ limit far more challenging than that of mesons, on the other hand allows one to apply a well-known method of
nuclear physics, the semiclassical mean-field theory. Indeed, as argued by Witten, baryons in the large $N_c$
limit contain $N_c$ quarks with n-quark force scaling as $N_{c}^{1-n}$. Thus, one can treat this weakly interacting many-body state
in a mean-field approximation. Unfortunately, the explicit treatment is only available for heavy non-relativistic quarks, in which case the
mean-field treatment corresponds to the Hartree approximation \cite{CohenKumarNdousse}. The picture that arises from such a treatment
is that of a baryon with a mass of order of $N_c$ and a size and shape which are independent of $N_{c}$.
Despite the fact that explicit mean-field treatment in the case of the light quarks is unknown, the large $N_c$ scaling for baryon
observables is expected to be valid.

Witten realized \cite{Witten} that the above scaling of mesons and baryons
indicates that the baryons in large $N_c$ limit
arise as quantized soliton-like configurations of mesonic fields.
A particular model which satisfies the large $N_c$ scaling is a well-known
Skyrme model \cite{Skyrme}. In this model the baryons appear as quantized skyrmions, the topological
solitons of a particular non-linear mesonic lagrangian\cite{ANW}. The stability of baryons as quantum
solitons is due to the existence of conserved topological current \cite{SkyrmeReview,SUNlattice}.

In addition to a single-baryon sector, it is of great interest
to consider baryon-baryon interaction in the large $N_c$ limit.
Since the baryon mass diverges in the large $N_c$ limit the
baryon-baryon scattering observables don't have a smooth limit
for scattering at fixed center-of-mass energy and momentum transfer.
For such momentum, $p \sim N_c$, one instead focuses on the potential
between two baryons which scales as $N_c$. In this context, the large $N_c$ scaling rules
had been used to analyze the spin-flavor structure of the nucleon-nucleon potential \cite{NNpotential}.
In particular, it was shown that at leading order the nucleon-nucleon potential is
symmetric under contracted $SU(4)$ spin-flavor symmetry \cite{SU2Nf}. In addition, one can also
address a question of consistency of the meson-exchange picture of the nucleon-nucleon
potential \cite{NNMesonExchange,CD}.

The focus of the present talk is on the kinematic regime corresponding to
a fixed center-of-mass velocity. In this case, both the kinetic and potential
energy of two baryons are of order $N_c$, and thus one expects a smooth limit
to exist for the scattering cross-section. As argued by Witten \cite{Witten}, here, as in the
case of the single-baryon sector, the appropriate framework is the mean-field
description. However, in this case one has to use {\it time dependent} mean-field
theory (TDMFT). As was shown in \cite{NNobservables}, one can discuss the spin-flavor structure of
the total nucleon-nucleon cross section
at leading order in large $N_c$ expansion. As in the case of the nucleon-nucleon potential,
the emergent contracted $SU(4)$ spin-flavor symmetry leads to certain relations between
total proton-proton and proton-neutron cross sections. It will be shown in section~\ref{sec:data} that
these
relations satisfy the behavior of the experimental nucleon-nucleon cross sections at the
center-of-mass energies of order of a few GeV.

\section{Time Dependent Mean Field Theory Framework}

The goal here is to show that time-dependent mean-field theory (TDMFT) framework valid in the
large $N_c$ limit, and contracted $SU(2N_f)$ symmetry, where $N_f$ is the number of light quark flavors,
to make model-independent predictions about the spin-isospin structure of the total nucleon-nucleon cross
sections. As discussed above, TDMFT treatment is a valid framework for the nucleon-nucleon scattering
when the center-of-mass transfer momentum is $p\sim N_c$. Since the nucleon size and hence the size of the
interaction region are of order of $N_{c}^{0}$, the scattering in this kinematic regime is semi-classical.

The description of interaction by TDMFT methods requires time-averaging over all field configurations
consistent with the initial state of two nucleons. This precludes one from being able to
calculate the S-matrix elements \cite{Griffin,SkyrmionSkyrmion}. However, as shown in \cite{NNobservables},
there are certain inclusive
nucleon-nucleon observables which can be evaluated in TDMFT framework. One such observable can be
formed from conserved baryon current whose expectation value in the initial two-nucleon state can
be in principle evaluated. The expectation values of the baryon current can be related to the inelastic
differential cross section.

In TDMFT framework each quark and gluon field of two nucleons move in a time-dependent field
created by all other quarks and gluons. These equations are not known explicitly. As a result, one can not
determine the nucleon-nucleon cross section even in the large $N_c$ limit. However, it is possible
to determine the spin-isospin dependence of the cross section using the contracted $SU(4)$ symmetry
valid in the large $N_c$ limit. Since the focus here on spin-isospin dependence of the
nucleon-nucleon scattering, one can use the Skyrme model which encapsulates the spin-flavor structure
of large $N_c$ baryons \cite{BaryonsModelIndependent}.

In the Skyrme model the nucleon dynamics is described in terms of classical soliton configurations built
out of pion fields. A convenient form for such a soliton is given by $SU(2)$-valued matrices
$U_h(\vec{r})=\exp \left(i\vec{\tau} \hat{n} F(r)\right)$, where $\tau^{a}$ are Pauli matrices,
$\hat{n}=\vec{r}/r$ and $F(r)$ is the magnitude of the pion field, $F(r)=|\vec{\pi}(r)|$ \cite{ANW}. Such
classical configurations impose correlations between spacial and isospin rotations and they are
referred to as {\it hedgehogs}. The baryons appear after the quantization of classical hedgehogs.
This is done by quantizing the slow rotation of the hedgehog in isospin space, $A^{\dag}(t)U_h(\vec{r})A(t)$,
given by the time-dependent $SU(2)$ matrix $A(t)$. These rotations describe the slow collective degrees
of freedom of the hedgehog, the zero-modes \cite{ZeroModes,ZeroModesTwo}.
After quantization, the generators of these vibrations proportional to
$Tr\left (\tau^{a}\dot{A(t)}A^{\dag}(t)\right)$ correspond to the spin and isospin quantum numbers of the
the ground-state band of states, $I=J=1/2, 3/2, ... N_c/2$ in the large $N_c$ limit.
The first two states correspond to nucleon and $\Delta$ baryons. The masses of these states are degenerate up to
the terms $I(I+1)/ M_B \sim 1/N_c$. This is a representation of the contracted spin-flavor symmetry $SU(4)$
in the context of the Skyrme model. The spin-isospin dependence of the wave-function of these states
is given by Wigner matrices $D^{I=J}_{m, m'} (A)$ where $m, m'$ are the third components of spin and isospin
respectively. Note that the Wigner matrices are functions of parameters of collective rotations $A$ and
{\do not} of $\dot{A}(t)$.

A crucial consequence of the above semi-classical analysis of the single-baryon sector in the Skyrme
model is the appearance of the scale separation in the dynamics of the collective degrees of freedom
described by the Wigner matrices $D^{I=J}_{m, m'} (A)$ at leading order in $1/N_c$ and intrinsic
degrees of freedom which describe all other non-collective excitations which include excited states of
the ground states baryons and mission of virtual and real mesons. The frequency of the collective
excitations are of order $1/N_c$ while that of the {\it intrinsic} excitations are of order $N_{c}^{0}$.
This scale separation enables an adiabatic or treatment of the collective degrees in the context of TDMFT
treatment of the nucleon-nucleon scattering analogous to the Born-Openheimer approximation in the context
of the rotational and vibrational excitations of molecules. There the slow degrees of freedom correspond
to vibrations of atomic nuclei in the averaged field produced by electrons whose motion represent
the intrinsic excitations.

To obtain an observable describing the nucleon-nucleon scattering one can start with
a function which describes the initial state of two well separated hedgehogs corresponding
to the initial state of two nucleons. As discussed in \cite{NNobservables}, in the context of the
Skyrme model it is convenient to choose a conserved baryon current
$J^{\mu} \left(\vec{r}, t; A_1, A_2, v, \vec{b}, \hat{n}\right)$.
In the Skyrme model it is a topological invariant.
The dependence of the current on the collective degrees of freedom are described by the variables,
$A_1, A_2, v, \vec{b}, \hat{n}$ where
$A_1, A_2, v, \vec{b}, \hat{n}$ define the spin and isospin configuration, the center-of-mass speed,
impact parameter and the unit vector along the direction separating the centers of the two hedgehogs
at the initial moment. The initial distance is not indicated. Such parametrization corresponds
to the semi-classical description of the scattering which as discussed above is valid in
the large $N_c$ limit. The functional dependence of this current can only be determined
once the explicit form of TDMFT equations is known. However, as shown below one does not
need to know these equations to determine the spin-isospin structure of the corresponding
scattering observable. This structure is determined by transformational properties of
the current under the spin and isospin rotations which is determined by the
contracted sin-flavor symmetry.

As shown in \cite{CohenGelman,NNobservables}, the classical current can be turned into a differential cross section by integrating
the current $J^{\mu} \left(\vec{r}, t; A_1, A_2, v, \vec{b}, \hat{n}\right)$
over time and the impact parameters,
\begin{equation}
\frac{d \sigma_{\rm inc}  (v,A_1,A_2;\theta,\phi)} {d\Omega} =
\lim_{R \rightarrow \infty} {R^2} \,
\int_0^\infty  db \, (2 \pi b)  \,
\int^{\infty}_{0} dt
\hat{r} (\theta,\phi) \cdot
\vec{J}\left(R\hat{r}(\Omega), t; A_1, A_2, v, \vec{b}, \hat{n}\right) \,,
\label{HHcrosssection}
\end{equation}
The above equations gives the probability for one
hedgehog to emerge in a cone with a solid angle $d\Omega$ around
a direction given by polar angles $\theta$ and $\phi$. The current
in Eq.~(\ref{HHcrosssection}) is normalized as to give the total
baryon number two. The time $t=0$ in Eq.~(\ref{HHcrosssection})
corresponds to the time at which two hedgehogs have the smallest
separation. The integral in Eq.~(\ref{HHcrosssection}) can be
explicitly evaluated only when TDMFT equations are known. It is also
important to note that the probability in Eq.~(\ref{HHcrosssection})
is integrated over all outgoing meson degrees of freedom. In the final
analysis it will give the inelastic cross section.

To turn the probability in Eq.~(\ref{HHcrosssection}) into
a nucleon-nucleon cross section one needs to evaluate
an expectation value of $d \sigma_{\rm inc}/d\Omega$
in an initial two-nucleon state described by
spin and isospin projections $m_{1}, m_{1}^{I}$ and
$m_{2}, m_{2}^{I}$ on the direction given by
the unit vector $\hat{n}$.
It can be done due to the scale separation between the collective
and intrinsic degrees of freedom discussed above. Indeed, the semiclassical
quantization of the baryon current in Eq.~(\ref{HHcrosssection}) leads
the appearance of the terms proportional to $J^2/M_N$ and $I^2/M_N$, where
$M_N$ is the nucleon mass. These terms represent coupling between the
collective and intrinsic excitations. However, as discussed above these
terms are of order $1/N_c$ and do not contribute at leading order in
the $1/N_c$ expansion. This result allows one to find expectation values
of the inclusive cross section using the superposition of the initial
hedgehogs described by the collective variables $A_1$ and $A_2$ weighted
by the corresponding Wigner D-matrix. In other words, the spin-isospin
part of the nucleon wave function in the initial state with given quantum numbers
$m, m^{I}$ is
\begin{equation}
|m^{J}, m^{I} > \, = \int dA \, \, D^{1/2}_{m, m^I} \,(A) \, |A > \,,
\label{NucleonWaveFunction}
\end{equation}
where $|A >$ represents a hedgehog with particular orientation
in spin-isopsin space, $D^{1/2}$ is the Wigner D-matrix describing
spin-isospin coordinates of the nucleon,
and the integral is taken over the space
of the collective coordinates.

Using Eq.~\ref{NucleonWaveFunction} one can find the inclusive
(integrated over all mesons in the final state) nucleon-nucleon
differential cross section at leading order in $1/N_c$ expansion,
\begin{equation}
\frac{d\sigma^{(m_1, m^I_1, m_2 , m^I_2)} (v, \theta,\phi)}{d\Omega} = \\
\int dA_1 dA_2 |D_{m_1,m^I_1}^{1/2}(A_1)|^2 |D_{m_2,m^I_2}^{1/2}(A_2)|^2 \,
\frac{d \sigma_{\rm inc}  (v,A_1,A_2;\theta,\phi)} {d\Omega} \,,
\label{sigmaInclusive}
\end{equation}
where the $d \sigma_{\rm inc}(v,A_1,A_2;\theta,\phi)/d\Omega$ is given in
Eq.~(\ref{HHcrosssection}). In Eq.~(\ref{sigmaInclusive}),
$m_1, m^I_1, m_2 , m^I_2$ are the spin and
isospin components of two nucleons along the direction $\hat{n}$ which
can be taken as the beam axis.

It is possible now to find a general form of the inclusive
differential cross section at leading order in $1/N_c$ by integrating
over the impact parameter space in Eq.~(\ref{HHcrosssection}) and
$SU(2)$ measure in Eq.~(\ref{sigmaInclusive}). The resulting expression found
in \cite{CohenGelman} is
\begin{eqnarray}
&d\sigma^{(m_1,m^I_1, m_2, m^I_2)} (v,\theta,\phi)/d\Omega= \nonumber \\
&a_0(v, \theta,\phi) + b_I(v, \theta,\phi) \, \left(\vec{\sigma}_1\cdot\vec{\sigma}_2\right)\,
\left(\vec{\tau}_1\cdot\vec{\tau}_2\right) +
c_I(v, \theta,\phi)\, \left(\vec{\sigma}_1\cdot\vec{n}\right)
\left(\vec{\sigma}_2\cdot\vec{n}\right)\,
\left(\vec{\tau}_1\cdot\vec{\tau}_2\right) \,,
\label{DSigma}
\end{eqnarray}
where $\sigma^{i}$ and $\tau^{a}$ are the spin and isospin Pauli matrices corresponding to
the two initial nucleons, and the functions $a_0$, $b_I$ and $c_I$ encode the leading order behavior at large $N_c$.
In obtaining the result in Eq.~\ref{DSigma} the following identity
of the Wigner $D$-matrices was used,
\begin{equation}
\left(D^{J}_{m, n}\right)^{*}= (-1)^{m-n} D^{J}_{-m,-n} \,.
\label{SpinIsospinFlip}
\end{equation}

Equation~(\ref{DSigma}) represents an inclusive differential cross section
at leading order in $1/N_c$ expansion. However, more readily available is the data
for total inelastic nucleon-nucleon cross section. To obtain
the total cross section one should integrate the differential cross section over
the whole solid angle. In doing so one obtains the following
expression,
\begin{equation}
\sigma^{(m_1,m^I_1, m_2 , m^I_2)} (v) =
A_0(v) + B_I(v) \, \left(\vec{\sigma}_1\cdot\vec{\sigma}_2\right)\,
\left(\vec{\tau}_1\cdot\vec{\tau}_2\right) +
C_I(v)\, \left(\vec{\sigma}_1\cdot\vec{n}\right)
\left(\vec{\sigma}_2\cdot\vec{n}\right)\,
\left(\vec{\tau}_1\cdot\vec{\tau}_2\right) \;,
\label{Sigma}
\end{equation}
where
$$
A_0 (v) =1/2\int d\Omega \,  a_0(v;\theta,\phi),  \,\,\,
B_I(v) =1/2\int d\Omega \,  b_I(v;\theta,\phi), \,\,\,
C_I(v) =1/2\int d\Omega \,  c_I(v;\theta, \phi).
$$
The factor of $1/2$ in the above equations are due to the normalization
of the baryon current.

However, while formally integrating over the solid angle to obtain Eq.~(\ref{Sigma})
we did not consider that as discussed above the TDMFT treatment from which the
differential cross section in Eq.~(\ref{sigmaInclusive}) was derived strictly holds
only in the semiclassical limit. As is well-known, \cite{L&L} the semiclassical approximation
breaks down for small forward angles for which $\theta \leq 1/R p_N$, where $R \sim N_{c}^{0}$
is a size of the interaction potential which is of the order of the nucleon size, and $p_N \sim N_c$
is the center-of-mass momentum. Thus, the forward angle at which the semiclassical approximation
breaks down is of order of $1/N_c$ in the kinematic region of interest here. However, the total
cross section in Eq.~(\ref{Sigma}) includes both forward and backward angles for which the semiclassical
approximation breaks down.

However, as shown in details in \cite{CohenGelman} the contribution to the total cross section from the forward angles
$\theta \leq 1/N_c$ vanishes as $1/N_c$ in the large $N_c$ limit provided the scattering cross section
is not anomalously peaked in a vanishingly small forward direction. The latter would require large elastic
contribution. Thus, the total cross section given in Eq.~(\ref{Sigma}) is valid up to corrections
of order $N_c$.

The key result of the above discussion is the form of total nucleon-nucleon cross section given in
Eq.~(\ref{Sigma}) which is valid up to corrections of order $1/N_c$. While the Skyrme model has been
used in the derivation, the result is independent of the details of the model and are valid in the
large $N_c$ limit. All the details of the dynamics for which the explicit TDMFT equations
have to be used are in the functions $A_0(v)$, $B_0(v)$ and $C_0(v)$ which are of order one in
the large $N_c$ limit. These functions can not be determined explicitly at present time. Nevertheless,
the form of the total cross section in Eq.~(\ref{Sigma}) does contain testable predictions. Note
that it is not the most general form which can be obtained based on parity and time reversal
invariance. For example, it does not contain such terms as
$A_I \left(\vec{\tau}_1\cdot\vec{\tau}_2\right)$ and
$B_0 \left(\vec{\sigma}_1\cdot\vec{\sigma}_2\right)$. These terms do not appear at leading
order in $1/N_c$. This result does not depend on the details of the Skyrme but follows
from the contracted spin-flavor symmetry $SU(4)$ applied to the two-nucleon sector.

This result can be tested against the existing experimental data for total nucleon-nucleon
cross sections. This is done in the next section \cite{CohenGelman}.

\section{\label{sec:data}Comparison with experimental data}

The experimental data exists for the total spin-independent and polarized proton-proton and
proton-neutron cross sections \cite{NNexperimental}. The kinematic regime in which the result obtained above is expected to
be valid corresponds to the center-of-mass momentum above $1\, GeV$.

\begin{figure}
\begin{center}
\includegraphics[width=0.9\textwidth]{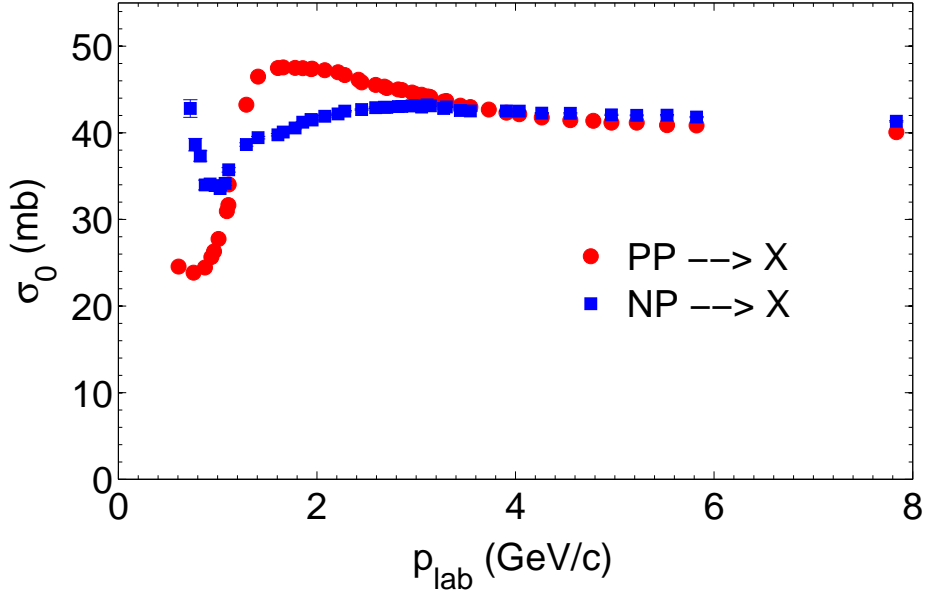}
\caption{Spin-averaged proton-proton and neutron-proton total cross section as a function
of beam momentum (Bugg et al., 1996).}
\label{SpinIndependentFig}
\end{center}
\end{figure}

The total nucleon-nucleon cross section given in Eq.~(\ref{Sigma}) contains the total nucleon-nucleon
cross sections for isosinglet and isotriplet initial states,
$$\sigma^{(m_1,m^I_1, m_2 , m^I_2)} (v) =
\left( \begin{array}{cc} \sigma^{(I=0)} & 0 \\ 0 & \sigma^{(I=1)} \end{array} \right)$$
Using projection operators $\left(1-\vec{\tau}_1\cdot\vec{\tau}_2\right)/4$ and
$\left(3-\vec{\tau}_1\cdot\vec{\tau}_2\right)/4$, one can extract $\sigma^{(I=0)}$ and
$\sigma^{(I=1)}$ cross sections. Then one obtains the total proton-proton,
$\sigma^{pp} = \sigma^{(I=1)}$,
and neutron-proton cross sections, $\sigma^{np} =  \frac{1}{2}\left(\sigma^{(I=1)} + \sigma^{(I=0)}\right)$.
Thus, at leading order in $1/N_c$ one has the following expressions,
\begin{eqnarray}
\sigma^{(pp)} & = &
A_0 + B_I \, \left(\vec{\sigma}_1\cdot\vec{\sigma}_2\right) +
C_I\, \left(\vec{\sigma}_1\cdot\vec{n}\right)
\left(\vec{\sigma}_2\cdot\vec{n}\right) \nonumber
\\
\sigma^{(np)} &=&
A_0 - B_I \, \left(\vec{\sigma}_1\cdot\vec{\sigma}_2\right)-
C_I\, \left(\vec{\sigma}_1\cdot\vec{n}\right)
\left(\vec{\sigma}_2\cdot\vec{n}\right) \,.
\label{channelSigmas}
\end{eqnarray}
Recall that in the above equations $\vec{n}$ represents the beam axis.
Using Eq.~(\ref{channelSigmas}) and averaging over the spin-polarization
of the beam and target nucleons, one can obtain spin-averaged total
cross sections. Thus, at leading order we have the following relation,
\begin{equation}
\sigma^{(pp)}_{0}= \sigma^{(np)}_{0} \left(1+{\cal{O}}(1/N_c)\right)\,,
\label{predictionSigma0}
\end{equation}
where $\sigma_0$'s are the spin-averaged total cross sections. This
prediction follows from the large-$N_c$ analysis and {\it cannot} be
obtained simply from isospin invariance.
This large-$N_c$ result is well satisfied by the experimental data
shown in Fig.~\ref{SpinIndependentFig}.

The total cross sections for the case when beam and target
nucleons are transversely polarized relative to the beam direction
can also be obtained. Two configurations
are possible, $\uparrow\uparrow$ and $\uparrow\downarrow$. These can be
combined into an observable,
$\Delta\sigma_T = -\left(\sigma(\uparrow\uparrow)- \sigma(\uparrow\downarrow)\right)$
referred to as {\it delta sigma transverse}.

Analogously, for the the longitudinally polarized beam and target nucleons
one can extract an observable, $\Delta\sigma_L = -\left(\sigma(\rightrightarrows)- \sigma(\rightleftarrows)\right)
$, the {\it delta sigma longitudinal}.

The large-$N_c$ analysis, Eq.~(\ref{channelSigmas}), predicts the following results for these observables,
\begin{eqnarray}
\Delta\sigma^{(pp)}_{T} & = & - \Delta\sigma^{(np)}_{T}
\left(1+{\cal{O}}(1/N_c)\right) \,, \nonumber \\
\Delta\sigma^{(pp)}_{L} & = & - \Delta\sigma^{(np)}_{L}
\left(1+{\cal{O}}(1/N_c)\right) \, .
\label{DeltaSigma}
\end{eqnarray}

\begin{figure}
\begin{center}
\subfigure[$\Delta\sigma_T$]{%
\label{DeltaSigmaTFig}
\includegraphics[width=0.5\textwidth]{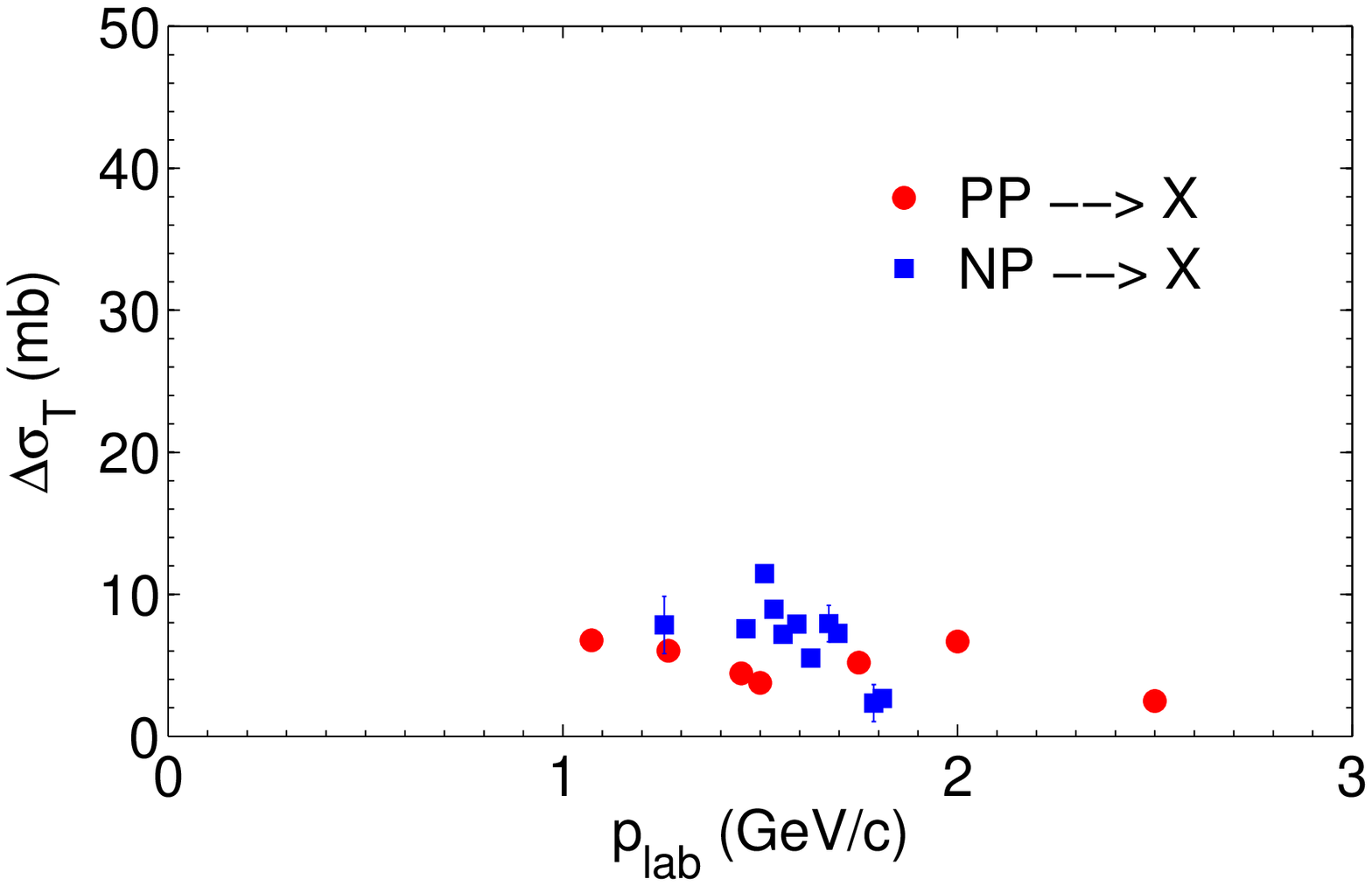}
}%
\subfigure[$\Delta\sigma_L$]{%
\label{DeltaSigmaLFig}
\includegraphics[width=0.5\textwidth]{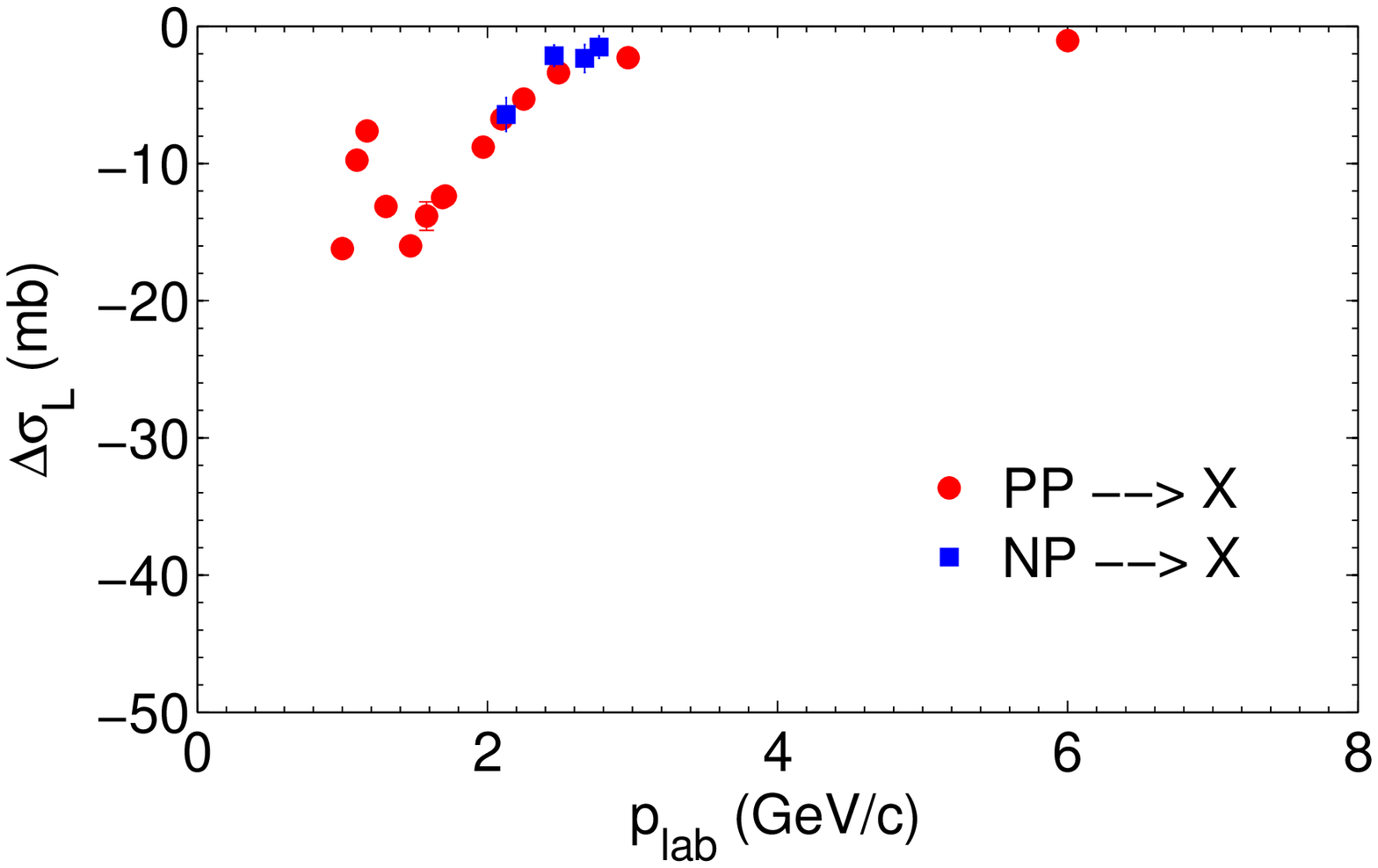}
}%
\end{center}
\caption{$\Delta\sigma_T$ for neutron-proton (Fonteine et al., 1991) and proton-proton
(Ditzler et al., 1983; Lesikar, J. D. 1981) scattering, and
$\Delta\sigma_L$ for proton-proton (Auer et al., 1978), and neutron-proton
(Sharov et al., 2008) scattering as functions of beam momentum.}
\end{figure}

Experimental data for these observables is shown in Figs.~\ref{DeltaSigmaTFig} and ~\ref{DeltaSigmaLFig}.
One may conclude from Figs.~\ref{DeltaSigmaTFig} and ~\ref{DeltaSigmaLFig} that the large $N_c$
results given in Eq.~(\ref{DeltaSigma}) are not satisfied by data. However, the results are valid within
corrections of order $N_c$. Indeed, according to Eq.~(\ref{Sigma}) both $\sigma_0$ and
$\Delta\sigma_T$ and $\Delta\sigma_L$ are of the same order in $1/N_c$. Experimentally however
$\Delta\sigma_T$ and $\Delta\sigma_L$ are much smaller then $\sigma_0$. The latter are about
$40\, mb$ while the former are consistent with zero. The suppression of $\Delta\sigma_T$ and $\Delta\sigma_L$
are for reasons not predicted by large-$N_c$ analysis. Nevertheless, qualitatively the predictions are
valid since $\Delta\sigma_T$ and $\Delta\sigma_L$ are small for both $pp$ and $np$ scattering.

\acknowledgments
BAG would like to gratefully acknowledge the support of
Professional Development Advisory Council of the New York City College of Technology,
the Professional Staff Congress-City University of New York Research Award Program through the grant PSCREG-41-540, and
the Center for Theoretical Physics, New York City College of Technology, The City University of New York.
The work on which this talk is based was done in collaboration with T.D Cohen.


\begin{thebibliography}{99}
\bibitem{CohenGelman} T. D. Cohen and B. A. Gelman, Phys. Rev. C {\bf 85} (2012) 024001.
\bibitem{NNobservables} T. D. Cohen and B. A. Gelman, Phys. Lett. B {\bf 540} (2002) 227.
\bibitem{tHooft} G. 't Hooft, Nucl. Phys. B {\bf 72} (1974) 461.
\bibitem{Witten} E. Witten, Nucl. Phys. B {\bf 160} (1979) 57.
\bibitem{CohenKumarNdousse} T. D. Cohen, N. Kumar and K. K. Ndousse, Phys. Rev. C {\bf 84} (2011), 015204.
\bibitem{Skyrme} T.H.R. Skyrme, Proc. Roy. Soc. A {\bf 260} (1961) 127.
\bibitem{ANW} G. S. Adkins, C. R. Nappi and E. Witten, Nucl. Phys. B {\bf 228} (1983) 552.
\bibitem{SkyrmeReview} I. Zahed and G. E. Brown, Phys. Rept. {\bf 142} (1986) 1.
\bibitem{SUNlattice} B. Lucini and M. Panero, arXiv:1210.4997 (2012).
\bibitem{NNpotential} D. B. Kaplan and M. J. Savage, Phys. Lett. B {\bf 365} (1996) 244;
D. B. Kaplan and A. V. Manohar, Phys. Rev. C {\bf 56} (1997) 76.
\bibitem{SU2Nf} J. L. Gervais and B. Sakita, Phys. Rev. Lett. {\bf 52} (1984) 87;
R. Dashen, E. Jenkins and A.V. Manohar, Phys. Rev. D {\bf 49} (1994) 4713;
C. Carone, H. Gergi and S. Osofsky, Phys. Rev. Lett. B {\bf 322} (1994) 227;
M. Luty and J. March-Russell, Nucl. Phys. B {\bf 426} (1994) 71.
\bibitem{NNMesonExchange}
M. K. Banerjee, T. D. Cohen and B. A. Gelman, Phys. Rev. C {\bf 65} (2002) 034011;
A. V. Belitsky and T.D. Cohen, Phys. Rev. C {\bf 65} (2002) 064008;
T. D. Cohen, Phys. Rev. C {\bf 66} (2002) 064003;
A. Calle Cordon and E. Ruiz Arriola, Phys. Rev. C {\bf 78} (2008) 054002; Phys. Rev. C {\bf 80} (2009) 014002.
\bibitem{CD} T. D. Cohen and D. C. Dakin, Phys. Rev. C {\bf 68} (2003) 017001.
\bibitem{Griffin} J. J. Griffin and M. Dworzecka, Phys. Rev. Lett B {\bf 93} (1980) 235.
\bibitem{SkyrmionSkyrmion} T.S. Walhout and J. Wambach, Phys. Rev. Lett. {\bf 67} (1991) 314;
T. Gisiger and M. B. Paranjape, Phys. Rev. D {\bf 50} (1994) 1010.
\bibitem{BaryonsModelIndependent} M. P. Mattis, Phys. Rev. D {\bf 39} (1989) 994;
E. Jenkins and R. F. Lebed, Phys. Rev. D {\bf 52} (1995) 282;
C.L. Schat, J. L. Goity and N. N. Scoccola, Phys. Rev. Lett. {\bf 88} (2002) 102002.
\bibitem{ZeroModes} S. Coleman, Aspects of Symmetry, Cambridge Univ. Press (1985).
\bibitem{ZeroModesTwo} R. Rajaraman, Solitons and Instantons: An Introduction to Solitons and Instantons in
Quantum Field Theory, North-Holland Physics Publishing, 1987.
\bibitem{L&L} L. D. Landau and E. M. Lifshitz, Quantum Mechanics: non-relativistic theory,
 3rd ed., Pergamon Press, 1977.

\bibitem{NNexperimental} C. Lechanoine-LeLuc and F. Lehar, Rev. Mod. Phys. {\bf 65} (1993) 47.




\end{thebibliography}
\end{document}